\documentclass{osa-article}
\journal{osajournal}

\begin{document}

\title{Mid-wave infrared super-resolution imaging based on compressive calibration and sampling}

\author{Xiao-Peng Jin,\authormark{1} Qing Zhao,\authormark{1,3,5} Xue-Feng Liu,\authormark{2,4,6} and An-Dong Xiong\authormark{1}}

\address{\authormark{1}Center for Quantum Technology Research and Key laboratory of Advanced Optoelectronic Quantum Architecture and Measurements (MOE), School of Physics, Beijing Institute of Technology, Beijing 100081, China\\
\authormark{2}Key Laboratory of Electronics and Information Technology for Space Systems, National Space Science Center, Chinese Academy of Sciences, Beijing 100049, China\\
\authormark{3}Beijing Academy of Quantum Information Sciences, Beijing 100193, China\\
\authormark{4}University of Chinese Academy of Sciences, Beijing 100049, China}

\email{\authormark{5}qzhaoyuping@bit.edu.cn}
\email{\authormark{6}liuxuefeng@nssc.ac.cn} 



\begin{abstract}
Mid-wave infrared (MWIR) cameras for large number pixels are extremely expensive compared with their counterparts in visible light, thus, super-resolution imaging (SRI) for MWIR by increasing imaging pixels has always been a research hotspot in recent years. Over the last decade, with the extensively investigation of the compressed sensing (CS) method, focal plane array (FPA) based compressive imaging in MWIR developed rapidly for SRI. This paper presents a long-distance super-resolution FPA compressive imaging in MWIR with improved calibration method and imaging effect. By the use of CS, we measure and calculate the calibration matrix of optical system efficiently and precisely, which improves the imaging contrast and signal-to-noise ratio (SNR) compared with previous work. We also achieved the $4 \times 4$ times super-resolution reconstruction of the long-distance objects which reaches the limit of the system design in our experiment.
\end{abstract}

\section{Introduction}
Mid-wave infrared (MWIR) is an invisible light range in the spectrum of 3 to 5 microns, compared with visible light. It has longer imaging distance, better ability to penetrate the smog and can work all-weather, thus has many applications in various fields \cite{2014Widely,2014Infrared,2006Measurement}. However, MWIR cameras for large number pixels are extremely expensive due to its native physical limitations. In many MWIR imaging system, the low number of imaging pixels becomes the critically restrictive factor of the imaging resolution instead of the diffraction limit as in visible imaging system. Therefore, how to increase the imaging pixels of MWIR optical system effectively has always been a key point for the researchers \cite{2008Teledyne,2009Large}.

Compressed sensing (CS) is proposed by Candes, Donoho and Tao in 2006. It proves one can recover the original sparse or compressible signal nearly identically only by a few non-adaptive linear measurements. Since the natural signal usually has sparse expression under a certain basis or framework, it can be effectively compressed during the projection process \cite{2004Robust,2006Compressed,2006Proceedings}. The CS theory provides a practical method to increase the imaging pixels of the system, that is, to reconstruct the high-resolution images from the low-resolution images measured by sensors. With the emergency of CS theory, many related studies are proposed. The single-pixel camera (SPC) was developed at Rice University in 2006 \cite{2006A}, then, based on which, Ke proposed the block-wise compressive imaging (CI) \cite{2012Object}, McMackin introduced the SPC in short-wave infrared in 2012 \cite{2012A}, A. Mahalanobis proposed the focal-plane array-based compressive imaging (FPA CI) in the mid-wave infrared in 2014 \cite{2014Recent}, Chen applied the FPA CI to the short-wave infrared \cite{2015FPA}, Wang summarized the MWIR FPA CI systematically in 2019 \cite{Zimu2019Focal}. In these works, high-resolution modulations are introduced to provide the final imaging resolution and the requirement of high-resolution FPA is relieved. However, it is worth noting that the studies above realized the super-resolution imaging mainly in the overall image size and qualitatively comparison, but there are comparatively few reports of super-resolution imaging which quantitatively evaluate the imaging resolution and truly reach the modulation level.

Optical system calibration is an important factor to affect the imaging quality and imaging resolution \cite{2016Computational}. As for infrared imaging, the physical properties and the experiment instruments cause the non-uniformity of the optical system \cite{Zimu2019Focal}, thus, calibration is inevitable for imaging process. The calibration-based methods like traditional point-scanning calibration approach are time-consuming and vulnerable to temperature instability, moreover, the low energy in a single point will reduce the contrast and signal-to-noise ratio of the calibration results and further decrease the imaging quality \cite{2014Recent,2011Calibration,1995Nonuniformity,2005New}. The scene-based calibration methods are inapplicable due to the greatly computational complexity caused by the system aberration \cite{2011Scene,2000Scene,1999Statistical}. The deep-learning based calibration methods require numerous datasets to estimate the parameter \cite{1991Adaptive}.

In this paper, to realize an accurate non-uniformity calibration of the compressive MWIR imaging system to guarantee the imaging resolution and quality, we firstly propose a non-uniformity calibration method based on CS theory which can calculate the calibration matrix efficiently and precisely against the negative influence of the low contrast and SNR in MWIR system. Then we construct a super-resolution compressive imaging approach, consisting of low-resolution measurement and high-resolution reconstruction, and design a MWIR FPA CI system focus on long-distance objects. Finally, by measuring and reconstructing the digital objects and physical objects, we verify the proposed approach effectively improves the imaging quality and fully utilizes the modulation resolution.

\section{Methods}
\begin{figure}[htbp]
	\centering
	\includegraphics[width=0.9\linewidth]{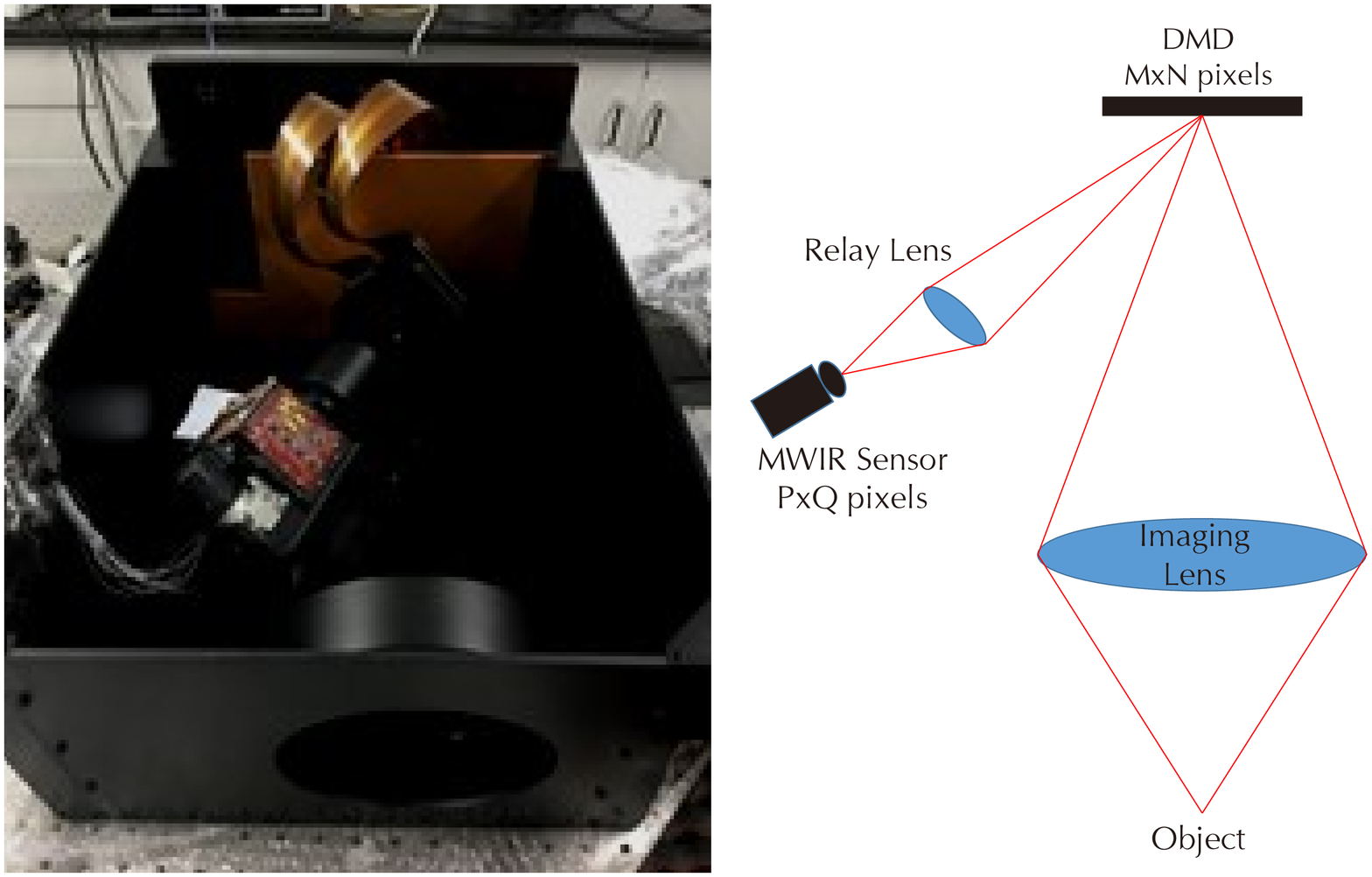}
	\caption{}
	\label{A}
\end{figure}

\noindent A traditional MWIR compressive imaging model is shown in Fig.~\ref{A}. The object is first imaged through the imaging lens on the digital micro-mirror device (DMD), which is an array formed by numerous micro-mirrors with each mirror being controlled by computer individually. Thus, we generated the modulation masks in advance, loaded them on the DMD and successively modulated the objects in high resolution. After that, the modulated high-resolution images reflected by DMD are projected onto the MWIR FPA sensor through the relay lens. Finally, the low-resolution images captured on FPA sensor are used for reconstruction of high-resolution original image by CS algorithm. The process above is normally expressed as $y = A\cdot x$, where $A \in {\mathbb{R}^{{M_1} \times {N_1}}}\left( {{M_1} \le {N_1}} \right)$ is the measurement basis or measurement matrix, $x \in {\mathbb{R}^{N_1}}$ represents the one-dimensional form image to be measured, $y \in{\mathbb{R}^{M_1}}$ is the coded compressive signal recorded by FPA sensor, and ${M_1}$ and ${N_1}$ are pixel numbers of detection and modulation, respectively. The reconstruction process is to solve an optimization problem of the following form:
\begin{equation}
 \hat x = \arg \min \frac{1}{2}\left\| {y - A \cdot x} \right\|_2^2 + \lambda \Psi \left( x \right),
\label{eq:refname1}
\end{equation}
where $\Psi \left( x \right)$ is the regularization term, $\lambda$ is a penalty parameter to balance the sparsity and the residual term. There are many CS algorithms for the reconstruction process, such as TVAL3 \cite{2011An}, TwIST \cite{2008A}, OMP \cite{1993Orthogonal}. In this paper, we apply TVAL3 for reconstruction, which can be written as the augment Lagrangian function: 
\begin{equation}
\mathop {\min }\limits_x \sum\limits_i {{{\left\| {{D_i}x} \right\|}_p}}  + \frac{\mu }{2}\left\| {y - A \cdot x} \right\|_2^2,
\label{eq:refname2}
\end{equation}
where ${\left\| x \right\|_p} = {\left( {\sum\nolimits_{i = 1}^N {{{\left| {{x_i}} \right|}^p}} } \right)^{\frac{1}{p}}}$, ${D_i}x$ denotes the discrete gradient vector of $x$ at the $i$th position, $\mu$ is a balance parameter and $p$ is the norm number. 

Next, we will firstly illustrate the optical calibration method based on CS, then describe the measurement and reconstruction process.
 
\subsection{Calibration Method}
In practice, the system non-uniformity caused by optical aberrations, distortion and vibration results in the imperfect under-sampling factor between the DMD pixels and sensor pixels, which denotes the number of DMD pixels mapped onto each sensor pixel \cite{2016Computational}. If an ideal under-sampling factor is used in the image reconstruction, the deviation from actual system will make the measurement matrix inaccurate and decrease the imaging quality. Therefore, before the experiment, the non-uniformity calibration for all the DMD and sensor pixels is inevitable.
\begin{figure}[htbp]
	\centering
	\includegraphics[width=0.9\linewidth]{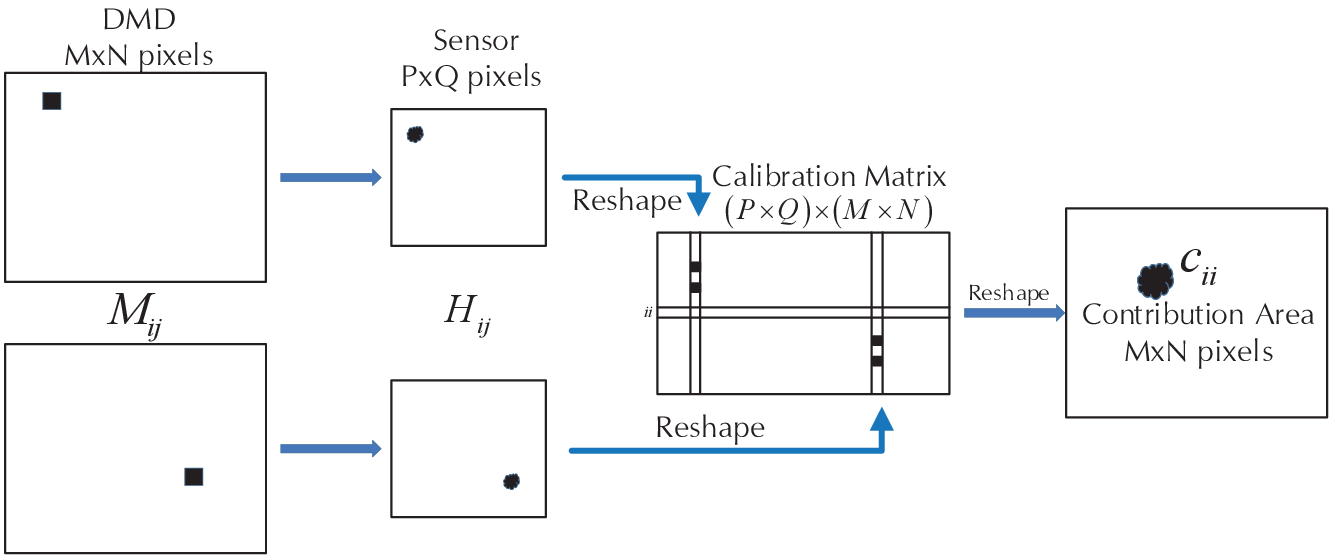}
	\caption{}
	\label{B}
\end{figure}

The typical calibration process is shown in Fig.~\ref{B}, with the size of DMD and FPA sensor be $M \times N$ and $P \times Q$. Beginning at the top left corner of the DMD, we successively display a sequence of matrixes ${M_{ij}}$ on DMD, the size of which are $M \times N$, for each time, the matrix has one pixel $\left( {i,j} \right)$ to be 1 and the rest are all $0$. Following this process, the corresponding images ${H_{ij}}$ with the size of $P \times Q$ are recorded on the sensor. We next reshape each ${H_{ij}}$ to a column and construct them to a new matrix $C$, which we term  as the calibration matrix. The size of calibration matrix is $\left( {P \times Q} \right) \times \left( {M \times N} \right)$, and we can get $L\left( {{H_{ij}}} \right) = C * L\left( {{M_{ij}}} \right)$, where $L\left(  \cdot  \right)$ denotes column-wise vectorization operator. It is worth noting that we reshape each row of the calibration matrix into a $M \times N$ matrix ${c_{ii}}$, $1 \le ii \le \left( {P \times Q} \right)$. We refer to this as the pixel contribution area which represents the contribution of each pixel on the DMD to a given sensor pixel $ii$ $\left( {1 \le ii \le \left( {P \times Q} \right)} \right)$.
The process above can construct the calibration matrix, however, in the actual experiment, this process is time-consuming. For instance, a DMD with size of $1920 \times 1280$ needs $1920 \times 1280 = 2457600$ times for measuring which is too inefficient. In addition, the luminous flux for each measurement is quite small, which is vulnerable to the system noise such as detector dark noise and environment temperature, resulting in the low contrast and SNR. To overcome these problems, we propose a high throughput, high efficiency calibration method based on CS. The details are as follows, we display $m$ frames of random binary masks $\left( {{{\rm{a}}_1},{{\rm{a}}_2},...,{{\rm{a}}_m}} \right)$ with the size of $M \times N$ sequentially onto the DMD and record the corresponding $m$ frames images $\left( {{{\rm{y}}_1},{{\rm{y}}_2},...,{{\rm{y}}_m}} \right)$ with the size of $P \times Q$ on the FPA sensor. Note that the infrared light of each sensor pixel is from its pixel contribution area, thus this process is actually the modulation for each pixel contribution area by random binary masks. For each FPA sensor pixel, we can obtain:  
\begin{equation}
\left[ {\begin{array}{*{20}{c}}
	{{y_{1,ii}}}\\
	{{y_{2,ii}}}\\
	\vdots \\
	{{y_{m,ii}}}
	\end{array}} \right] = \left[ {\begin{array}{*{20}{c}}
	{{{\rm{a}}_1} \odot {c_{ii}}}\\
	{{{\rm{a}}_2} \odot {c_{ii}}}\\
	\vdots \\
	{{{\rm{a}}_m} \odot {c_{ii}}}
	\end{array}} \right],\left( {1 \le ii \le \left( {P \times Q} \right)} \right)
\label{eq:refname3}
\end{equation}
where $\odot$ denotes inner product, ${{y_{m,ii}}}$ is the value of $ii$th pixel in ${y_m}$. To make this a mathematically formula, we have:
\begin{equation}
\left[ {\begin{array}{*{20}{c}}
	{{y_{1,ii}}}\\
	{{y_{2,ii}}}\\
	\vdots \\
	{{y_{m,ii}}}
	\end{array}} \right] = \left[ {\begin{array}{*{20}{c}}
	{{{\left( {L\left( {{{\rm{a}}_1}} \right)} \right)}^T} \cdot L\left( {{c_{ii}}} \right)}\\
	{{{\left( {L\left( {{{\rm{a}}_2}} \right)} \right)}^T} \cdot L\left( {{c_{ii}}} \right)}\\
	\vdots \\
	{{{\left( {L\left( {{{\rm{a}}_m}} \right)} \right)}^T} \cdot L\left( {{c_{ii}}} \right)}
	\end{array}} \right] = \left[ {\begin{array}{*{20}{c}}
	{{{\left( {L\left( {{{\rm{a}}_1}} \right)} \right)}^T}}\\
	{{{\left( {L\left( {{{\rm{a}}_2}} \right)} \right)}^T}}\\
	\vdots \\
	{{{\left( {L\left( {{{\rm{a}}_m}} \right)} \right)}^T}}
	\end{array}} \right] \cdot L\left( {{c_{ii}}} \right) \Rightarrow y = a \cdot c,
\label{eq:refname4}
\end{equation}
$1 \le ii \le \left( {P \times Q} \right)$, where $T$ denotes the transpose operator, $a$ is a $m \times \left( {M \times N} \right)$ matrix, $c$ is a $\left( {M \times N} \right) \times 1$ matrix, and $y$ is a $m \times 1$ matrix. After reconstructing each $c$ by TVAL3, we get each pixel contribution area ${c_{ii}}$ and the calibration matrix eventually. By this method, light from approximately half pixels of DMD in average is collected to the sensor. Compared with only one pixel in the traditional calibration, the flux for each measurement is greatly increased, and the contrast and SNR of the captured images can be improved. In addition, because of the sparsity of the calibration matrix and the sub-sampling ability of CS, the modulation and measurement times can be much less than the image pixels. Therefore, the whole measurement process is efficient, which can also reduce the influence of the system noise.

\subsection{Measurement And Reconstruction}
In the typical FPA CI, researchers always divide the DMD area according to the theoretical under-sampling factor, which results in a group of SPCs in parallel. However, such operation is not suitable in our imaging system. The system non-uniformity leads to the imperfect projections between DMD and sensor pixels against the theoretical under-sampling factor, therefore, for each single-pixel camera, part of the measured signal comes from DMD pixels outside the theoretical corresponding area. Besides, after the individual reconstruction for each SPC, the value difference between pixels on each boundary will cause the sharp stripe in the final image, which reduces the quality of image. As a consequence, the reconstruction of the whole image based on the calibration matrix is feasible in our system instead of a group of independent SPCs. 

Our imaging system is shown in Fig.~\ref{A}. We choose the Hadamard16 matrix for sampling as the system theoretical under-sampling factor is approximate $\left( {4 \times 4} \right):1$ (described in next section). In the measurement process, we firstly change the “-1” elements of the Hadamard16 to “0” in order to match the two states of DMD, and then reshape each row of the revised Hadamard16 to a ${\rm{4}} \times {\rm{4}}$ matrix to get 16 basis matrix. Next each basis matrix is expanded to an overall ${M \times N}$ matrix $\left( {Mas{k_1},Mas{k_2}, \ldots ,Mas{k_{16}}} \right)$ as the masks to be displayed onto the DMD. The object image is finally modulated by the masks and captured by FPA sensor as the corresponding ${P \times Q}$ patterns $\left( {Imag{e_1},Imag{e_2}, \ldots ,Imag{e_{16}}} \right)$. To make this process mathematically concrete, we get $L\left( {Imag{e_i}} \right) = C \cdot diag\left( {Mas{k_i}} \right) \cdot x$, where $x$ denotes the column-wise vectorized version of the ${M \times N}$ object image, $L\left( {Imag{e_i}} \right)$ denotes the column-wise vectorized version of the captured images on sensor, $diag$ means forming a diagonal matrix, with elements in the diagonal line be that of column-wise vectorized version of the former matrix. As the result, after measurement, we get:
\begin{equation}
\left[ {\begin{array}{*{20}{c}}
	{L\left( {Imag{e_1}} \right)}\\
	{L\left( {Imag{e_2}} \right)}\\
	\vdots \\
	{L\left( {Imag{e_t}} \right)}
	\end{array}} \right] = \left[ {\begin{array}{*{20}{c}}
	{C \cdot diag\left( {Mas{k_1}} \right) \cdot x}\\
	{C \cdot diag\left( {Mas{k_2}} \right) \cdot x}\\
	\vdots \\
	{C \cdot diag\left( {Mas{k_t}} \right) \cdot x}
	\end{array}} \right] = \left[ {\begin{array}{*{20}{c}}
	{C \cdot diag\left( {Mas{k_1}} \right)}\\
	{C \cdot diag\left( {Mas{k_2}} \right)}\\
	\vdots \\
	{C \cdot diag\left( {Mas{k_t}} \right)}
	\end{array}} \right] \cdot x \Rightarrow I = A \cdot x,
\label{eq:refname5}
\end{equation}
$\left( {1 \le t \le 16} \right)$, where $x$ is a $\left( {M \times N} \right) \times 1$ column vector, $I$ is a $\left( {P \times Q \times t} \right) \times 1$ column vector and $A$ is a $\left( {P \times Q \times t} \right) \times \left( {M \times N} \right)$ matrix. Afterwards, the TVAL3 is applied to reconstruct the original high-resolution object image in our experiment.

\section{System Description}
\begin{figure}[htbp]
	\centering
	\includegraphics[width=0.7\linewidth]{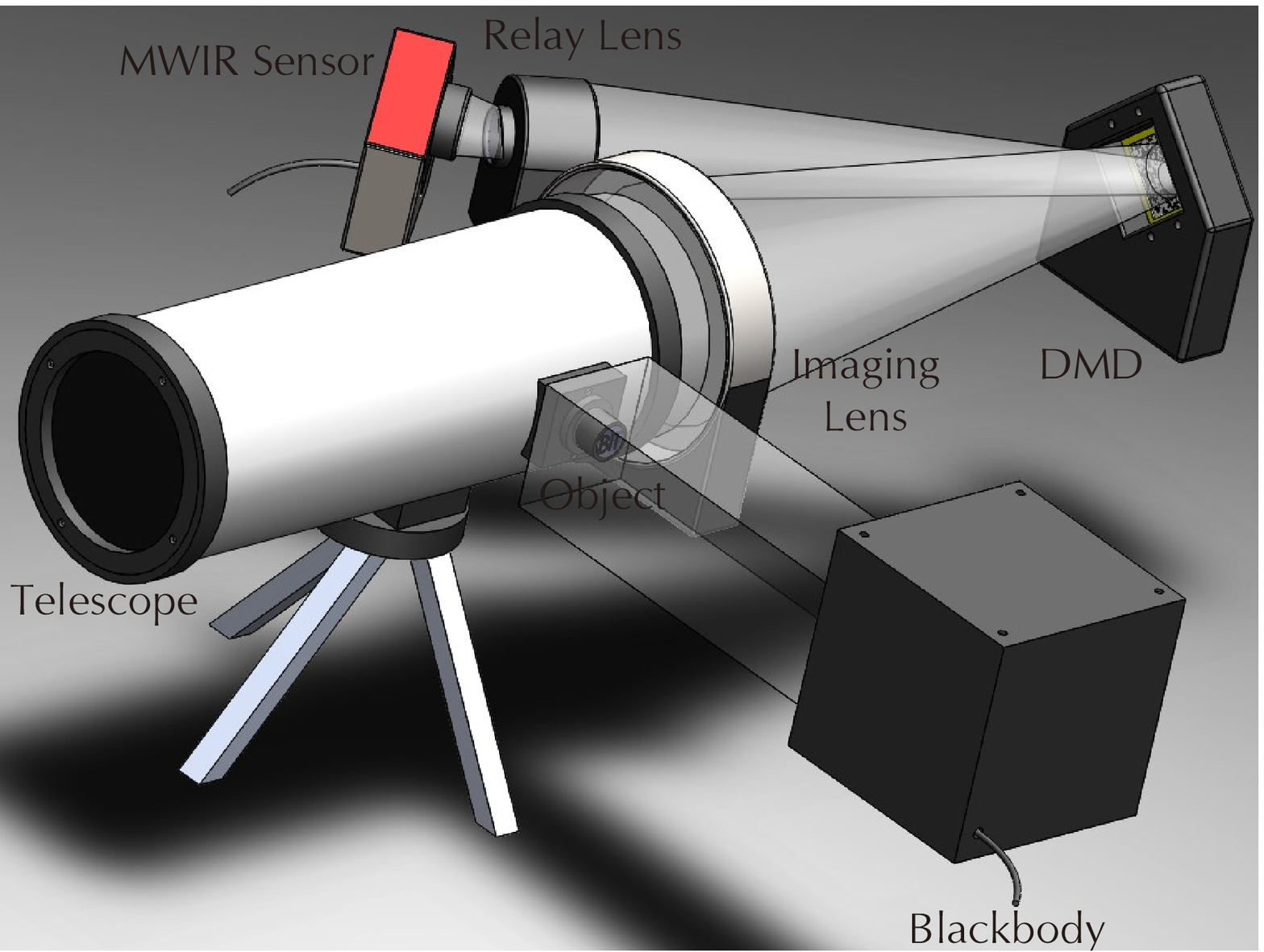}
	\caption{}
	\label{C}
\end{figure}

\noindent In the MWIR compressive imaging system as shown in Fig. 1, the object is imaged on the DMD through an imaging lens with 300mm focal length. The resolution of DMD is $1920 \times 1280$ (DLP9500, TI, USA), each micro-mirror in which is $10.8\mu m$ and can rotate to either ${\rm{ + 12}}$ or ${\rm{ - 12}}$ degree independently. We replaced the original window of DMD by sapphire for the high transmittance in the MWIR. The high-resolution images on DMD are captured by a MWIR sensor (Sofradir, IRE-320) which resolution is $320 \times 256$ with response spectral range of 3.7-4.8$\mu m$ through a specially designed relay lens. In addition, the cooling instruments are placed on DMD and sensor to reduce the system noise. In our actual experiment, the effective sizes of DMD and sensor are $1280 \times 1024$ and $320 \times 256$ respectively, thus, the system theoretical under-sampling factor is $\left( {4 \times 4} \right):1$.

Furthermore, as the experiment system is designed for the long-distance object, the depth of field for the telescopic imaging lens is relatively small, which demands the DMD and imaging lens be placed parallel such that the object can be clearly imaged on the DMD. As a result, the non-parallel between DMD and sensor causes the optical aberration, which leads to two defects. One is the deformation after imaging against the original object, and the other is the focus area for object becomes smaller. Therefore, we adjust the position of the relay lens and the FPA sensor to satisfy the Scheimpflug principle for a larger clear imaging area \cite{2006Programmable}.

To verify the proposed method and simulate the long-distance object precisely, we add some instruments to established a test system as shown in Fig. 3. The infrared source is a planar type blackbody with temperature ranging in 5 to 300$^{\circ}$C (IRSV, HT-20-D150), and the telescope system is a reflection type astronomical telescope (BOSMA750150) with 750mm focal length. The object is placed on the view port of the telescope, the infrared light transmitting from which will be shaped into parallel light to simulate an object at infinity.

\section{Experiment And Discussion}

In our actual experiment, a group of $2 \times 2$ DMD mirrors called a super-pixel is used. The reason is the diffraction limited spatial resolution of the relay imaging system is about $17\mu m$ according to calculation, and the size of each DMD mirror is $10.8\mu m$. Therefore, this group of mirrors is better matched to the sizes above. Besides, as in the conventional optical system design it is usually demanded that the size of imaging pixel be half of the expected imaging resolution, $2 \times 2$ DMD mirrors will be the minimum size to be distinguished in the compressive imaging system. Unless otherwise noted, the following experiments are all operated under the notation of super-pixel. In addition, to evaluate the image quality, the peak signal-to-noise ratio (PSNR) is introduced for quantitative analysis:
\begin{equation}
PSNR = 10\log \left( {{{{{\left( {{2^n} - 1} \right)}^2}} \mathord{\left/
			{\vphantom {{{{\left( {{2^n} - 1} \right)}^2}} {MSE}}} \right.
			\kern-\nulldelimiterspace} {MSE}}} \right),
MSE = \frac{1}{{MN}}\sum\nolimits_{i,j = 1}^{M,N} {{{\left[ {\tilde X\left( {i,j} \right) - X\left( {i,j} \right)} \right]}^2}}, 		
\label{eq:refname6}
\end{equation}
where $n$ is the bit number of sensor pixel value which is 14 in our experiment, and MSE describes the squared distance between the reconstructed image ${\tilde X}$ and the original image $X$. In general, the PSNR is positively correlated to the quality of image. In the following, all the images are normalized for comparison.

\subsection{Non-uniformity Calibration}

We firstly verify the non-uniformity calibration method proposed above. For comparison, the traditional point-scanning calibration method is applied in the $64 \times 64$ DMD area with the blackbody temperature of 200$^{\circ}$C. Due to the image deformation caused by optical aberration, we successively capture 4096 images in the sensor with the size of $20 \times 20$ pixels rather than $16 \times 16$ according to the system theoretical under-sampling factor of $\left( {4 \times 4} \right):1$. Then we get a $400 \times 4096$ traditional calibration matrix. The largest effective region of the DMD pixel contribution area for one pixel in the FPA sensor is about $8 \times 8$, thus, we display $m = 100$ frames of random binary masks for unified measurement and reconstruction. The results are shown in Figure 4. Fig. 4(a) represents the sum version of pixel-contribution areas of randomly selected 6 pixels based on the traditional calibration method. Fig. 4(b) shows the same version based on the proposed calibration method. It could be clearly seen that the pixel contribution areas from both methods are roughly identical, but has differences in exact pixel values and background. We think there are several reasons for this result. First, from the Fig. 4(a), the background noise is comparatively obvious, which results in the low contrast and SNR, while in our proposed method the 100 random masks increase the total flux for each measurement and improve the contrast and SNR of the captured images. Second, the sparsity reconstruction of CS algorithm can resist the measurement noise and give a result with very low background, which can be obviously seen in Fig. 4(b). Finally, the great amount of measurements in the traditional method is time-consuming and vulnerable to the system noise, which is also harmful to the calibration effect.

\begin{figure}[htbp]
	\centering
	\includegraphics[width=0.9\linewidth]{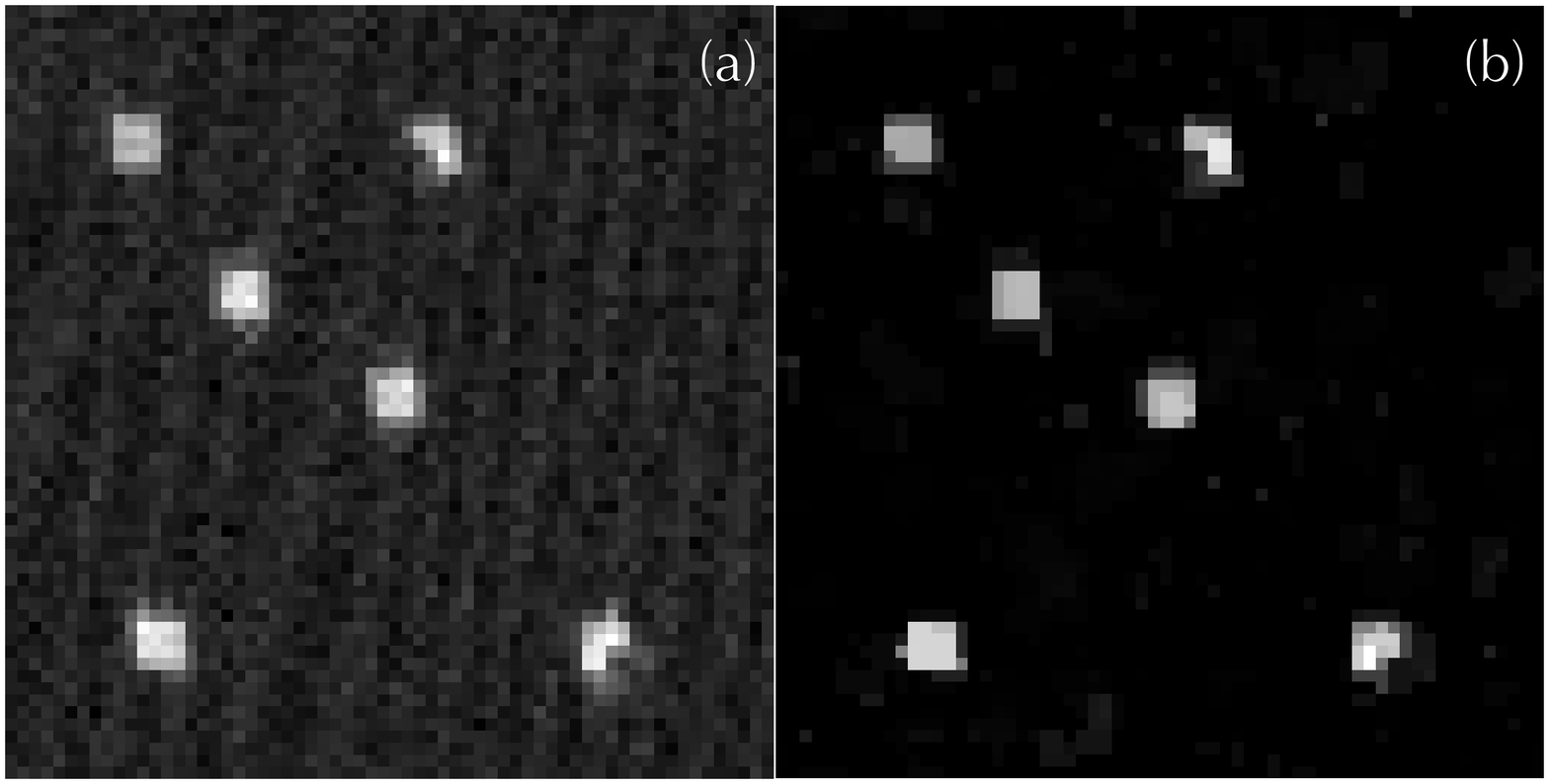}
	\caption{}
	\label{D}
\end{figure}

\subsection{Reconstruction Of Digital Object}

We next measure and reconstruct the digital object with two methods. The digital resolution target is loaded on the $64 \times 64$ DMD region with the blackbody temperature of 200$^{\circ}$C. The widths of fringes are 1, 2 and 4 pixels, respectively, as shown in Fig. 5(a). The values of pixels in the fringes are 1, and the background is 0. According to our method, the sampling ratio is:
\begin{equation}
ratio = n * \frac{{size\left( {FPA} \right)}}{{size\left( {DMD} \right)}},
\label{eq:refname7}
\end{equation}
where $n$ is the modulation mask number, ${size\left( {FPA} \right)}$ is the low-resolution image size captured by the sensor, and ${size\left( {DMD} \right)}$ is the high-resolution modulated image size displayed on the DMD. Due to the system theoretical under-sampling ratio is $\left( {4 \times 4} \right):1$, the formula above is actually: $ratio = {n \mathord{\left/
		{\vphantom {n {16}}} \right.
		\kern-\nulldelimiterspace} {16}}$. The reconstruction results and details for digital object based on two different former calibration matrices are shown in Figure 5. Fig. 5(b) is a direct observation of the original object by the sensor without a coded mask, which size is $20 \times 20$ pixels. Fig. 5(c) shows the pixel values along the vertical yellow dotted line in Fig. 5(b). Figs. 5(d-f) show the point-scanning calibration matrix based reconstructions of the object with the modulation mask number of 6, 10, 16 respectively. The small image placed in the right corner is the enlarged details of the red box. Each line profile in Fig. 5(h-j) is a vertical slice taken at the yellow dotted line from the corresponding upper images in Figs. 5(d-f). Figs. 5(k-n) and Figs. 5(p-r) are the reconstructions based on the proposed calibration matrix. The sampling ratios are the same as in Fig. 5(d-f).
	
In contrast to the low-resolution image in Fig. 5(b) and line profile in Fig. 5(c) which is indistinguishable of the one-pixel slits, the proposed method can clearly illustrate more details in the one-pixel slits in Figs. 5(k-m). Besides, by comparing Fig. 5(c) with Fig. 5(p), we prove the $4 \times 4$ times resolution enhancement of the proposed method, which means the high modulation resolution is fully utilized. Further, the recovered images based on the proposed method have higher qualities under the same or even lower sampling ratio against the traditional method through the PSNR value. This benefits from the more accurate calibration matrix measurement with the proposed method. Specifically, we increase the infrared flux during the calibration matrix measurement which improve the contrast and SNR and apply the CS algorithm for the better reconstruction. In addition, we greatly reduce the time for sampling, thus decrease the negative influence of the system noise.

\begin{figure}[htbp]
	\centering
	\includegraphics[width=0.8\linewidth]{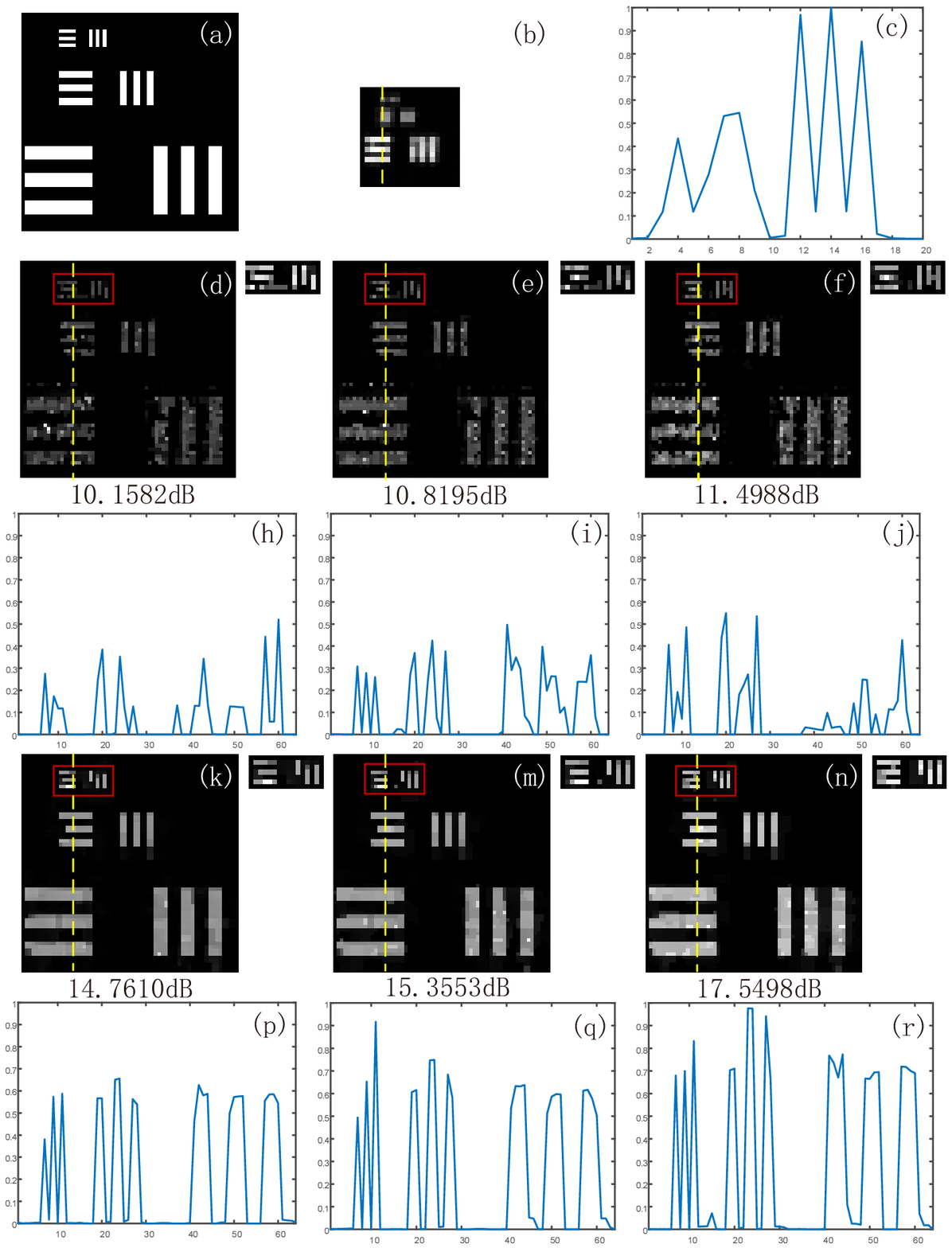}
	\caption{}
	\label{E}
\end{figure}

\subsection{Reconstruction Of Physical Object}

\begin{figure}[htbp]
	\centering
	\includegraphics[width=0.9\linewidth]{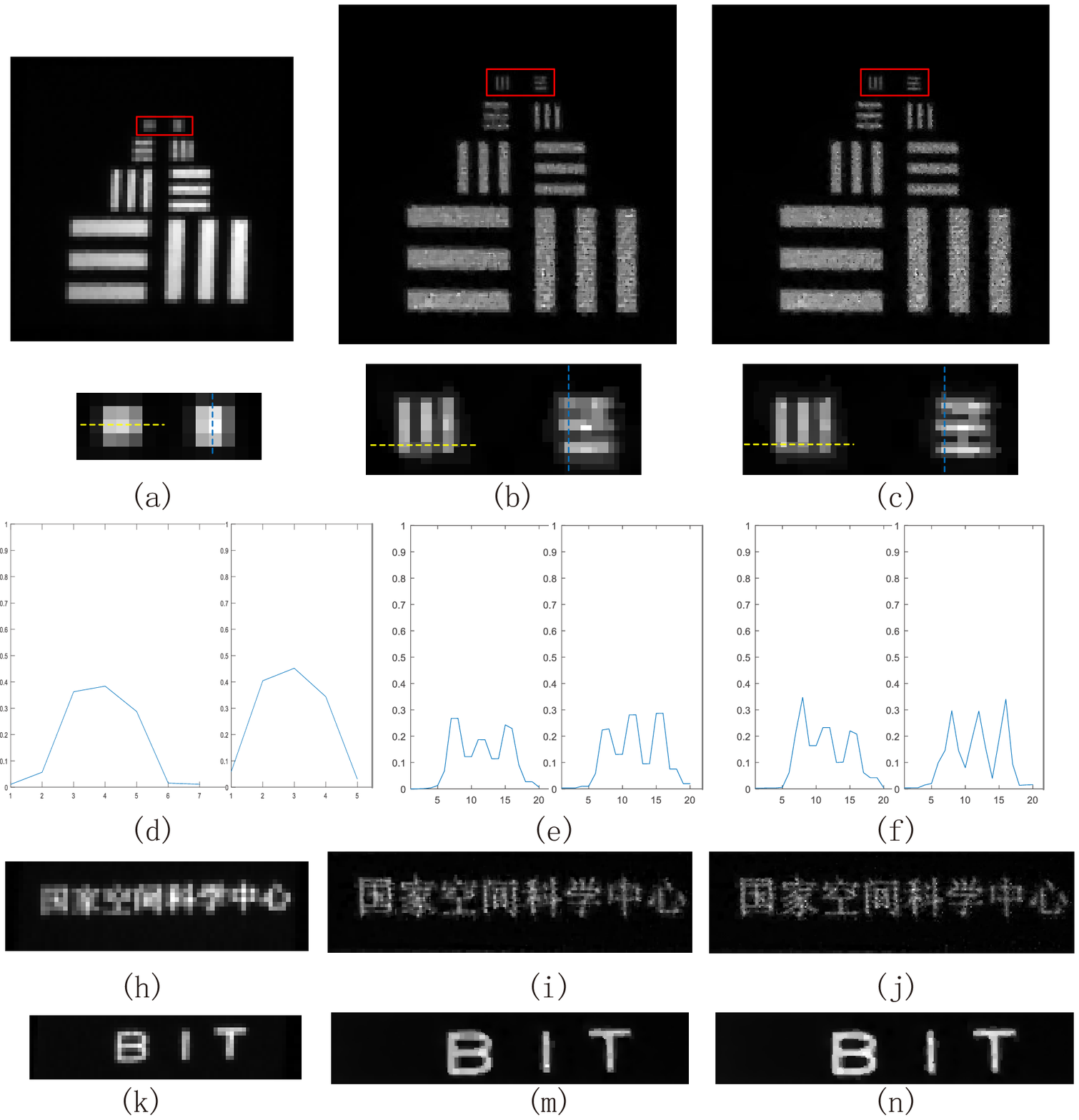}
	\caption{}
	\label{F}
\end{figure} 
\noindent Finally, we use the optical architecture to image some physical objects with the blackbody temperature of 200$^{\circ}$C, a resolution target, the letters of “BIT” (abbreviation of Beijing Institute of Technology) and the Chinese characters of “National Space Science Center”, which are placed in the view port of the telescope respectively. After the illumination, the object is projected to $256 \times 256$ pixels of DMD region and $68 \times 68$ pixels on the sensor. As the effectiveness of the proposed method is already verified in previous section, we display 100 random masks to calculate the calibration matrix. It should be emphasized that because the DMD pixel contribution area is limited to a small area of sensor, the calibration matrix for the larger region can be divided to several regions and reconstructed independently. Therefore, 100 masks are enough for the calibration at any pixel scale. Then we image the objects with 6 and 16 Hadamard masks and subsequently reconstruct the original object. It is worth noting that the 16 Hadamard masks are expanded to fit the DMD size by the basis hadamard16 in section 2.2. Limited by the manufacture accuracy, the smallest stripe in the resolution target is $75\mu m$, which is corresponding to 2 pixels on the DMD after projection. Fig. 6(a) shows the low resolution measurement of $68 \times 68$ pixels with an all-one modulation mask. Figs. 6(b-c) are the reconstructions of $256 \times 256$ pixels corresponding to 6 and 16 masks, and the images below are the enlarge details for the red boxes with the size of $5 \times 14$, $20 \times 50$, $20 \times 50$ pixels, respectively. Figs. 6(d-f) shows the pixel values along the horizontal yellow dotted lines and the vertical blue dotted lines in Figs. 6(a-c). The proposed method can recover the deformed object perfectly which can be seen in the bottom left corner of the Fig. 6(a) and Fig. 6(b). Further, for the indistinguishable stripes in Fig. 6(a), we can achieve the super-resolution reconstruction and illustrate superior image details which indicates the feasibility and effectiveness of the proposed method. Figs. 6(h-j) and Figs. 6(k-n) are the low-resolution images and high-resolution reconstructions for the Chinese characters and “BIT” letters. The pixel sizes of Fig. 6 (h), (i-j), (k), (m-n) are $20 \times 68$, $71 \times 256$, $16 \times 68$ and $20 \times 256$, respectively. It is clearly that through the compressive imaging system the imaging resolutions of different types of objects can all be improved.

\section{Conclusion}

In this paper, we have described the theory and experiment for the long-distance MWIR pixel super-resolution imaging, established the non-uniformity calibration method and image reconstruction based on CS. Through much fewer measurements compared with traditional point-scanning method, the non-uniformity calibration can be achieved more rapidly and precisely based on the subsampling and high-flux properties of CS. We also achieve the super-resolution reconstruction from a small number of low-resolution measurements. It is proved that using the accurate calibration matrix obtained by the proposed method, the super-resolution reconstruction quality can be effectively improved. Guaranteed by this, the high-resolution modulations are fully utilized and $4 \times 4$ times enhancement in the actual imaging resolution is realized.

In future work, we will focus on three main aspects. First, we will apply this imaging system to measure the actual long-distance object, which will face more serious impact on environment noise through the propagation. Second, as the computational speed in the reconstruction is restricted by the imaging pixel numbers, we will improve the efficiency of the reconstruction algorithm. Finally, the super-pixel is used to fit the diffraction limit of our system, thus, we try to propose a method to achieve the super-resolution imaging beyond the diffraction limit.

\section*{Funding}
National Key Research and Development Program of China (Grant no.2018YFB0504302).

\section*{Disclosures}
The authors declare no conflicts of interest.

\bibliography{sample}

\end{document}